\documentstyle[11pt,fleqn]{article}

\begin{document}
\baselineskip=18pt
\begin{center}
{\Large \bf Factoring the Unitary Evolution Operator and Quantifying
Entanglement}\\
\ \ \\
\ \ \\
X.X. Yi$^{1,2}$,C.P.Sun$^2$\\
{ $^1$\it Institute of Theoretical Physics, Northeast Normal University, \\
Changchun
130024, China}\\
{$^2$\it Institute of Theoretical Physics, Academia Sinica, Peking 100080,
China}\\
\end{center}
\vskip 1cm
{\bf Abstract}\\
The unitary evolution can be represented by a finite product of exponential
operators. It leads to a perturbative expression of the density operator of
a close system. Based on the perturbative expression scheme, we present a
entanglement measure, this measure has the advantage that it is easy to
compute for a general dynamical process.
\\
{\bf PACS numbers:89.70.-w,89.80.+h,03.65.Bz}\\
E-mail:llwls@ivy.nenu.edu.cn
\newpage
\section{Introduction}
\setcounter{equation}{0}
Over the past decade information theory has been generalized to include
quantum mechanical systems, for example, a two-level
quantum system has come to be known as a qubit in this context.
The additional freedom introduced with the quantum mechanical superposition
principle has opened up a variety of capabilities that
go well beyond those of conventional information techniques. There are
two distinct directions in which progress is currently being made:
quantum computation and error correction or prevention
on the one hand[1], and nonlocality
and distillation, on the other hand[2]. In each of those progresses,
quantum entanglement that provides a good measure of quantum correlations
plays an important role.

There are a number of good measures of the amount of entanglement for two
quantum systems in a pure state, a good measure of entanglement for mixed
states is also found though it is hard to compute for a general state[3].

In present paper,
considering the entanglement from the other aspect,
we  prefer to discuss the change of entanglement due to
the state changes rather than to compute straightforwardly
the entanglement of an arbitrary
state.

In the framework of quantum information theory, the state change
allowed by quantum mechanics may by treated in terms of quantum
operations[3], a simple example is  the unitary evolution experienced by a
close quantum system. The final state of the system is related to the
initial state by a unitary transformation $U$,
$$\rho\rightarrow \varepsilon(\rho)=U\rho U^+.$$
Unitary evolution is widely in use of quantum gates and circuits[4] as
a quantum operation. In addition to the unitary evolution, environment
coupling to a quantum system or a measurement performed on the quantum system
changes the state too[6,7]. The connection of quantum operations to
quantum measurements is easy to explain. Standard text book treatments
describe quantum measurement in terms of a complete set of orthogonal
projection operators for the system being measured. This formalism, however,
does not describe many of the measurements that can be performed on a
quantum system. The most general type of measurement that can be performed on
a quantum system is known as generalized measurement[6]. Generalized
measurement can be understood within the framework of unitary evolution,
because most generalized measurements can be realized through many dynamical
processes[8], and the state change due to environment may be also
treated in terms
of the unitary evolution, since an arbitrary open system may be enlarged by
including the environment to be a close system. In this sense, unitary evolution
is one of the most general types of state change possible in quantum mechanics.

The rest of present paper is organized as follows:
In Sec.2, we present a general method to factorize the unitary
evolution  operator $U(t)$ for a close system. The results may be generalized
in the treatment of many autonomous dynamical systems. Sec.3 contains our
results on the entanglement change occurring in a dynamical process. Finally, in Sec.4,
we present two typical  examples and some conclusions.

\section{Factorizing the unitary evolution operator $U(t)$}
\setcounter{equation}{0}

As noted above, the unitary evolution operator is one of the most general
types of state change possible in quantum mechanics, the point of this
section is to factorize the evolution operator into a
set of independent one. For this end, we discuss the following cases.\\
{\it case A[9]---} The Hamiltonian can be written as a finite sum
\begin{equation}
H(t)=\sum_i^m a_i(t)H_i,
\end{equation}
where $a_i(t)$ are a set of linearly independent complex valued functions
of time, and $H_i$ are constant operators. In addition, the set of operators
$H_i(i=1,...,m)$ may be enlarged by repeated commutation to a Lie algebra $L$
with finite dimension $n(n\geq m)$. With this presupposition, the unitary evolution
operator can be uncoupled into a set of independent operators
\begin{equation}
U(t)=U_1(t)U_2(t)...U_n(t),
\end{equation}
where each component $U_i(t)$ is an operator satisfying
\begin{equation}
\frac {d}{dt}U_i(t)=\dot{g}_i(t)H_iU_i(t), U_i(0)=1.
\end{equation}
With the scalar function $g_i(t)$ being the solution to a set of nonlinear
differential equations
\begin{equation}
\frac{d}{dt}g_i(t)=\sum_{i=1}^{n}\eta_{ik}a_k(t), g_i(0)=0,
\end{equation}
where $\eta_{ik}$ are nonlinear function of $g$'s. Thus we have factorized the
unitary evolution operator into the form:
\begin{equation}
U(t)=\prod_{i=1}^{n}e^{g_i(t)H_i}.
\end{equation}
Especially, for a  general case of a dynamically closed quantum system which
consists of two interacting subsystems $A$ and $B$,  the total
Hamiltonian may be written as a sum of three terms
\begin{equation}
H=H_A+H_B+H_{int},
\end{equation}
the first two terms represent the free Hamiltonian of subsystem $A$ and $B$,
respectively, and the last term describes the interaction between
the $A$ and $B$. Following the procedure stated above, we arrive at
\begin{equation}
U(t)=\prod_{i=AB,int,...,M} e^{g_i(t)}H_i.
\end{equation}
Here, $H_i(i=A,B,int,...M)$ are elements of Lie algebra with
finite dimension enlarged by
$H_A, H_B, H_{int}$
.\\
{\it case B---} In the case of the dimension of the Lie algebra enlarged by
$\{ H_i\}$ is infinite, we can factorized the unitary evolution operator
using the general Baker-campbell-Hausdorff[10] formula.
To start with, we give the evolution operator of the system
under consideration
\begin{equation}
U(t)=e^{-itH}=e^{-it(H_A+H_B+H_{int})},
\end{equation}
where $H_A,H_B,H_{int}$ are the same as in eq.(2.6), the eq.(2.8)
can approximately be written as
\begin{equation}
U(t)=e^{-\frac 1 2 [H_A+H_B,H_{int}]t^2}\cdot e^{-iH_{int}t}\cdot
e^{-i(H_A+H_B)t}+O(t^3).
\end{equation}
This splitting formula is hold in the case that $t$ has to be safely
smaller than a typical energy of the system. Thus, even in the simplest case,
a better method is needed. Let $n$ be a positive integer. The exponential
function satisfies the scaling identity
\begin{equation}
exp(-iHt)=[exp(-iHt/2^n)]^{2n}.
\end{equation}
When $n$  is sufficiently large, the argument $\frac {t}{2^n}$ is in some
sense small. Eqs.(2.9) and (2.10) together give
\begin{eqnarray}
U(t)&=&e^{-\frac 1 2 [H_A+H_B,H_{int}]\tau ^2}\cdot e^{-iH_{int}\tau }\cdot
e^{-i(H_A+H_B)\tau }e^{-\frac 1 2 [H_A+H_B,H_{int}]\tau ^2}\cdot \nonumber\\
&\cdot &e^{-iH_{int}\tau }\cdot
e^{-i(H_A+H_B)\tau }...e^{-\frac 1 2 [H_A+H_B,H_{int}]\tau ^2}\cdot
e^{-iH_{int}\tau }\cdot
e^{-i(H_A+H_B)\tau }+O(\tau ^3),
\end{eqnarray}
where $\tau=\frac{t}{2^n}$.
Still higher-order formulae are known.
We would like to point out that, in quantum computation[11], the $n=1$
is widely taken in use and it is large enough to
avoid the decoherence during quantum computing.

\section{Quantification of entanglement}
\setcounter{equation}{0}

In the previous section we have factorized the
time evolution operator $U(t)$. The question remains open
about how does the entanglement change in a dynamical process.
Of course,
this question is not entirely well defined unless
we state what physical circumstances characterized the amount of entanglement.
This suggest that there is no unique measure of entanglement.
Before we define the measure of entanglement we expand the density operator
for a close system. Suppose that the two interacting subsystems are initially
separable[12], i.e., their initial density operator (state) can be written in
a form
\begin{equation}
\rho(0)=\rho_A(0)\otimes\rho_B(0),
\end{equation}
to use the entanglement for quantum information processing, however, we need
a inseparable state, more precisely, a state in pure entanglement form.
The procedure of converting a separable state to inseparable one can be
performed, as stated in Sec.1, through a unitary evolution operator
$U(t)$(in addition, a partial trace is also needed sometimes.)
\begin{equation}
\rho(t)=U(t)\rho(0)U^+(t).
\end{equation}
If the interaction between the two subsystems is small, it is natural to
attempt some sort of Taylor series expansion of the
exponential in eqs. (3.2) (2.8) and (2.11), which give\\
{\it case A}
\begin{eqnarray}
\rho(\lambda,t)&=&\rho_A^0(t)\otimes\rho_B^0(t)+\lambda\sum_{i\neq A, B}
(\frac{\partial f_i}{\partial \lambda}H_i\rho_A^0(t)\otimes\rho_B^0(t)+
\rho_A^0(t)\otimes\rho_B^0(t)
\frac{\partial f_i^*}{\partial\lambda}H_i)\nonumber\\
&+&\frac{\lambda^2}{2}\sum_{i,j\neq A,B}(\frac{\partial f_i}{\partial\lambda}
\frac{\partial f_j}{\partial \lambda}H_iH_j\rho_A^0(t)\otimes\rho_B^0(t)
+\rho_A^0(t)\otimes\rho_B^0(t)\frac{\partial f_i}{\partial\lambda}
\frac{\partial f_j}{\partial \lambda}H_iH_j\nonumber\\
&+&\frac{\partial f_i}{\partial\lambda}H_i\rho_A^0(t)\otimes\rho_B^0(t)\frac
{\partial f^*_j}{\partial\lambda}H_j)+O(\lambda^2)
\end{eqnarray}
and
{\it case B}
\begin{eqnarray}
\rho(\lambda, t)&=&\rho_A^0(t)\otimes\rho_B^0(t)-\frac{\lambda}{2}
(\frac{t}{2^n})^2\sum_{i=0}^{2^n-1}\{ [H_A+H_B,H_{int}(t_i)]_-,
\rho_A^0(t)\otimes\rho_B^0(t)\}_+\nonumber\\
&-&i\lambda(\frac{t}{2^n})\sum_{i=0}^
{2^n-1}[H_{int}(t_i),\rho_A^0(t)\otimes\rho_B^0(t)]_-+O(\lambda^2)+
O((\frac{t}{2^n})^3),
\end{eqnarray}
where $\lambda$ denotes the coupling constant, and $\rho_i^0(t)$ represents
the state of subsystem $i$ at time $t$ with $\lambda=0$.
The results presented above suggest that we may take the form
\begin{equation}
\delta D(\rho)=||\rho(t)-\rho_A^0(t)\otimes\rho_B^0(t)||^2=
Tr(\rho(t)-\rho_A^0(t)\otimes\rho_B^0(t))^2
\end{equation}
as a measure of entanglement change in the time evolution process.
Noticing the initial state is separable,
the measure of entanglement change (3.5) is a measure of entanglement
in reality.
Although the definition of the measure for entanglement is not unique,
it has to satisfies the three condition stated below[3]:\\
(1)$D(\rho)=0$ if and only if $\rho$ is separable.\\
(2)Local unitary operators leave $D(\rho)$ invariant, i.e.
$$D(\rho)=D(U_A\otimes U_B\rho U_A^+\otimes U_B^+).$$
(3)The expected entanglement cannot increase under
Local general measurements+Classical communication+Postselection
(LGM+CC+PS) given by
$\sum_iV_i^+V_i=1,$i.e.,
$$\sum Tr(\rho_i)D(\rho_i/Tr\rho_i)\leq D(\rho),$$
where$\rho_i=V_i\rho V_i^+.$
For the measure of entanglement change proposed above, (1)follows from the fact
that $D(\rho)$ is a true metric, and (2) is obvious. Property (3)
is satisfied too[3].
We believe that there are numerous other nontrivial choices for measure
of entanglement, one of the choices could not be said to be more important
than any other,the present
choice has the advantage that it is easy to compute for
any dynamical process.

 Our discussion so far has centered on the entanglement change
in a dynamical process. To complete it we still need to show that this
definition can be generalized for any process that quantum mechanics
allowed. For a general process, the quantum operator that change a state
of the system should be factorized by the subsystem's operators.
For instance, a control not operation in quantum computation given by
(in fact, control not is a unitary evolution operator)
\begin{equation}
O=|0\rangle_1 \langle 0|\otimes 1_2+|1\rangle_1  \langle 1|\otimes
\sigma_2^x,
\end{equation}
where $1_2$ is the unit operator for the second qubit, $\sigma_2^x$
stands for the $x$ pauli matrix of the second qubit. $|1\rangle_1$
and $|0\rangle_1$ represent two state of the first qubit. This control
not operator is factorisable, i.e. $O$ can be written in the form
$$O=\sum_i O_1^i\otimes O_2^i,$$
where $O_1^i$ and $O_2^i$ denote operators for the first and second qubit,
respectively. Hence, according to the definition eq.(3.5), the control not
operator in the form (3.6) does not change the entanglement of the system.

We would like to point out that the discussions presented here are for the
unitary evolutions, for non-unitary evolution such as a trace over some
of the degree of freedom, we should find a auxiliary unitary process instead
of the non-unitary one.

\section{Example and Conclusion}
\setcounter{equation}{0}

In order to understand how our program for calculating the amount
of entanglement change works, we present in this section two examples,
one of them consists of two interacting qubits
(two identical two-level system) in a laser beam[5a],
and the another two independent qubits coupling simultaneously to a bath.\\
{\it example 1:}\\
The Hamiltonian describing the system in this example
has the following
form(set $\hbar=1$):
\begin{eqnarray}
H&=&H_A+H_B+H_{int}+H_f,\nonumber\\
H_i&=&\frac 1 2 \omega\sigma_z^i,(i=A,B)\nonumber\\
H_{int}&=&g(\sigma_A^+\sigma_B^-+\sigma_A^-\sigma_B^+)
+\lambda(\sum_{i=A,B}\sigma_i^+a+\sigma_i^-a^+),\nonumber\\
H_f&=&\omega_fa^+a,
\end{eqnarray}
where $\sigma_i^z,\sigma_i^-,\sigma_i^+$ describe the pauli
operator of the $i$ qubit, $g$ denotes the coupling constant,
and $H_f$ stands for the free Hamiltonian of the laser beam.
Suppose the state is initially in the form
\begin{eqnarray}
\rho(0)&=&\rho_A(0)\otimes\rho_B(0)\otimes\rho_f(0),\nonumber\\
\rho_A(0)\otimes\rho_B(0)&=&
|e_A,e_B\rangle\langle e_A,e_B|,\nonumber\\
\rho_f(0)&=&\sum_n p(n)|n\rangle\langle n|,
\end{eqnarray}
where $|e_i\rangle$ denotes the excited state of the qubit $i$
and $|n\rangle$ stands for a Fock state of the laser beam field.
In the Schr\"odinger
picture, the density operator that obeys von Neumann equation is given by

$$
\rho_{AB}(t)=Tr_f\rho(t)=
$$
{\tiny
$$
\left (
\begin{array}{lll}
\sum_np^2(n)f^2_{gg}(n,t) & \sum_np(n+1)p(n)f_{gg}(n+1,t)f_{EG}(n,t) &
\sum_np(n+2)p(n)f_{ee}(n+2)f_{gg}(n) \\
\sum_np(n+1)p(n)f_{gg}(n+1,t)f_{EG}(n,t) &\sum_n p^2(n)f_{EG}^2(n,t) &
\sum_n p(n)p(n+1)f_{EG}(n+1)f_{ee}(n)\\
\sum_np(n+2)p(n)f_{ee}(n+2)f_{gg}(n) & \sum_n p(n)p(n+1)f_{EG}(n+1)f_{ee}(n)
&\sum_n p^2(n)f_{ee}^2(n,t)
\end{array}
\right ),
$$
}
\begin{equation}
\ \
\end{equation}
where we take $|g_A,g_B\rangle, |E,G\rangle,$ and $|e_A,e_B\rangle$ as
a set of basis, and
$$|g_A,g_B\rangle=|g_A\rangle\otimes|g_B\rangle, |e_A,e_B\rangle=
|e_A\rangle\otimes|e_B\rangle,|E,G\rangle=\frac{1}{\sqrt{2}}
(|g_A\rangle\otimes|e_B\rangle+|e_A\rangle\otimes|g_B\rangle),$$
$$\rho_A^0(t)\otimes\rho^0_B(t)=\rho^0_A(0)\otimes\rho^0_B(0),$$
$$f_{gg}(n,t)=\frac 1 4\sin 2\phi\sin\theta e^{-iE_+t}+\frac 1 2
\sin 2\phi \cos^2\frac{\theta}{2}e^{-iE_-t}-\frac 1 2 \sin 2\phi
e^{-iE_0t},$$
$$f_{ee}(n,t)=\sin^2\phi\sin^2\frac{\theta}{2}e^{-iE_+t}+\cos^2
\phi e^{-iE_0t}+\sin^2\phi\cos^2\frac{\theta}{2}e^{-iE_-t},$$
$$f_{EG}(n,t)=\sin \theta \sin \phi \sin \frac{\Omega t}{2}, $$
$$E_{\pm}=\Omega\frac{\cos\theta\pm 1}{2}+\omega_f(n+1),E_0=(n+1)\omega_f,$$
and$ \theta=\frac{\pi}{2},\tan\phi=\sqrt{\frac{n+2}{n+1}},\Omega^2=
(16n+24)g^2.$
Eqs.(3.5) and (4.3) together give
\begin{equation}
\delta D(\rho)=\sum_{i,j=1,2,3}(\rho_{AB}^{ij})^2-2\rho_{
AB}^{33}+1,
\end{equation}
where $\rho_{AB}^{ij}$ denotes the element of matrix $\rho_{AB}$ given
by eq.(4.3), which represents
the entanglement change or entanglement of subsystems
$A$ and $B$ at time $t$. \\
{\it example 2}\\
The Hamiltonian describing dissipation of the two qubits has the following
form[1d](setting $\hbar=1$)
\begin{eqnarray}
H&=&\omega_0(\sigma_a^z+\sigma_b^z)+\sum_{l=a,b}\int d\omega [g_{\omega l}
A_{ l}(a^+_{\omega l}+a_{\omega l})]\nonumber\\
&+&\int d\omega \cup_{l=a,b}(\omega a_{\omega l}^+a_{\omega l}),
\end{eqnarray}
where $\vec{\sigma}_i$ describe the pauli's matrix of the $i$ qubit,
$a_{\omega l}$ stands for the bath mode $\omega$ coupling to the $l$
qubit, and $\cup_{l=a,b}a_{\omega l}^+a_{\omega l}=a_{\omega l}^+a_{\omega l}$
for $a_{\omega a}=a_{\omega b}$, whereas
$\cup_{l=a,b}a_{\omega l}^+a_{\omega l}=a_{\omega a}^+a_{\omega a}
+a_{\omega b}
^+a_{\omega b}$
for $a_{\omega a}\neq a_{\omega b}$. The coupling coefficients are denoted by
$g_{\omega l}$, and the qubit operator $A_l$ in general is expressed as
a linear superposition of three pauli's operators, i.e.
$A_l=\lambda^{(1)}\sigma_l^x+\lambda^{(2)}\sigma_l^y+\lambda^{(3)}\sigma_l^z$.
The ratio $\lambda^{(1)}:\lambda^{(2)}:\lambda^{(3)}$ is determined by the
type of the dissipation. For instance, $\lambda^{(1)}=\lambda^{(2)}=0$ for
phase damping and $\lambda^{(3)}=0$ for amplitude damping[7]. Phase
damping induces pure dephasing, whereas amplitude damping induces loss
and dephasing simultaneously. Many source of decoherence in quantum computers
are described by amplitude damping[13].

Without any loss of generality, we discuss in detail
the case with $\lambda^{(1)}
=\lambda^{(2)}=0$ i.e. phase damping.
Some words of caution are now in order. As mentioned above, the bath
can also cause another unwanted effect in computation process, i.e. amplitude
dissipation. It is easy, however, to make system have small loss rate
of amplitude dissipation[11], so a considerable number of operations are
allowed to perform.

In the case of phase damping, the unitary evolution operator may be
factorized
in the following form
\begin{equation}
U(t)=U_0(t)U_I(t),
\end{equation}
where $U_0(t)=e^{-iH_0t}$ with $H_0=
\omega_0(\sigma_a^z+\sigma_b^z)+
\int d\omega \cup_{l=a,b}(\omega a_{\omega l}^+a_{\omega l})$ is the
free evolution operator, while $U_I(t)$ denotes the evolution operator in the
interaction picture. A readily calculation shows that
\begin{eqnarray}
U_I(t)=U_I^a(t)U_I^b(t),\nonumber\\
U_I^i(t)=u_I^i(t)|e_i><e_i|+v_I^i|g_i><g_i|,
\end{eqnarray}
where $|e_i>$ and $|g_i>$ are the eigenstates of $\sigma_i^z$ with eigenvalues
$+1$ and $-1$, respectively, and $u_I^i$ and $v_I^i$ satisfy($i=a,b$):
\begin{eqnarray}
i\hbar\frac{\partial}{\partial t}u_I^i&=&\sum d\omega g_{\omega i}
(a_{\omega i}^+e^{i\omega t}+a_{\omega i}e^{-i\omega t})u_I^i,\nonumber\\
i\hbar\frac{\partial}{\partial t}v_I^i&=&-\sum d\omega g_{\omega i}
(a_{\omega i}^+e^{i\omega t}+a_{\omega i}e^{-i\omega t})v_I^i.
\end{eqnarray}
The Wei-Norman's algebraic method [9] that provides a way to factorize
the evolution operator gives
\begin{equation}
u_I^i=\prod_{\omega}e^{f_{\omega}^i(t)}e^{A_{\omega}^i(t)a_{\omega}^+}
e^{B_{\omega}^i(t)a_{\omega}}
\end{equation}
and
\begin{equation}
v_I^i=\prod_{\omega}e^{h_{\omega}^i(t)}e^{C_{\omega}^i(t)a_{\omega}^+}
e^{D_{\omega}^i(t)a_{\omega}}.
\end{equation}
Here,
$$A_{\omega}^i(t)=-\frac{g_{\omega i}}{\omega}(e^{i\omega t}-1),$$
$$B_{\omega}^i(t)=-(A_{\omega}^{i}(t))^*,$$
$$f_{\omega}^i(t)=-i\frac{g_{\omega i}^2}{\omega}t+\frac{g_{\omega i}^2
}{\omega^2}(1-e^{-i\omega t}),$$
$$C_{\omega}^i(t)=-A_{\omega}^i(t), D_{\omega}^i(t)=-B_{\omega}^i(t),
h_{\omega}^i(t)=f_{\omega}^i(t).$$
Now, we turn our attention to compute the reduced density operator of
the two-qubit system, first of all, we calculate the total density operator,
which follows straightforwardly from eq.(4.6)
\begin{equation}
\rho(t)=U_0U_I\rho(0)U_I^+U_0^+,
\end{equation}
where $\rho(0)$ denotes the initial density operator (state),
which may be written in a separable form
$$\rho(0)=\rho_a(0)\otimes\rho_b(0)\otimes\rho_B(0).$$
Here, $\rho_i(0)(i=a,b)$
represents the initial state of qubit $i$, and $\rho_B(0)$
stands for the initial state of the bath.
In following,
we use the notation,$|e_a,g_b\rangle$
to indicate the eigenstates of $\sigma_a^z$ and
$\sigma_b^z$ with eigenvalues $1$ and $-1$,  $Tr_B$ indicate a
trace over the bath, and $\rho^0_i(t)$ to represent
the free two-qubit state i.e.
$$\rho^0(t)=Tr_BU_0(t)\rho(0)U^+_0(t).$$
With this notation,
in a subspace spanned by
$\{|11\rangle=|e_a,e_b\rangle,|12\rangle=|e_a,g_b\rangle,|21\rangle=
|g_a,b_2\rangle,|22\rangle=|g_a,g_b\rangle \}$, the state of the
two-qubit system at time t takes the form
\begin{equation}
\rho_{ab}(t)=Tr_B\rho(t)=
\left ( \begin{array}{lccr}
\rho_{1111}&  \rho_{1112}&\rho_{1121}&\rho_{1122}\\
\rho_{1211}&\rho_{1212}&\rho_{1221}&\rho_{1222}\\
\rho_{2111}&\rho_{2112}&\rho_{2121}&\rho_{2122}\\
\rho_{2211}&\rho_{2212}&\rho_{2221}&\rho_{2222}
\end{array}
\right ) ,
\end{equation}
where
$\rho_{ijkl}=\rho_{ijkl}^0\cdot F_{ijkl}(i,j,k,l =1,2)$,
$\rho_{ijkl}^0=$ $Tr_B\langle ij|\rho^0(t)|kl\rangle,$ and
$F_{ijkl}=Tr_B$ $\langle ij|\sum_{c,d,e,f=1}^2(U_I^a)_i $ $
(U_I^b)_j|cd\rangle$ $\langle ef|
((U_I^b)^+)_k((U_I^a)^+)_l|kl\rangle.$
Here, $(U_I^a)_i=_a\langle i|U_I^a|i\rangle_a, $ and $ |2\rangle_a=|e_a\rangle$,
$|1\rangle_a=|g_a\rangle.$
In order to get more information about the reduced density
operator, we make some discussion on the quantity $F_{ijkl}$. It
can be easily verified that $F_{ijkl}=1$ for $i=k , j=l$,
while $F_{ijkl}=F_{klij}^*$
for $i\neq k$ and $j\neq l$.
Moreover, the quantity results from the interaction between
the two-qubit system and the bath, hence it depends on the states of the
bath. Although different bathes result in different results $F_{ijkl}$, the
physical results discussed here do not rely on the bath. In
this sense, we may consider a simple case  with zero temperature. In this
case, $F_{ijkl}$ is given that
\begin{equation}
F_{ijkl}=F_{ijkl}(t)=e^{-\int_0^{\infty}[\Delta_{ik}(\omega,t)
+\Delta_{jl}^*(\omega,t)]\rho(\omega)d\omega},
\end{equation}
where $\Delta_{ij}(\omega,t)=2\frac{(g_{\omega i}-g_{\omega j})^2
\sin^20.5\omega t}{\omega^2},$
$\rho(\omega)$ stands for the bath spectrum distribution.
The eq.(4.13) suggests that $F_{ijkl}$
approaches zero with the passage of time except
some moments at which
$\int_0^{\infty}[\Delta_{ij}(\omega,t)\Delta^*_{kl}(\omega,t)
]\rho(\omega)d\omega=0.$
This  attractive results might be used in preventing
information loss stored in quantum states. Now
we come back to the entanglement change, eqs. (3.5) and (4.12)
together give
\begin{equation}
\delta D(\rho_{ab}(t)||\rho_a^0(t)\otimes\rho_b^0(t))=\sum_{i,j,k,l=1}^2
(\rho_{ijkl}-\rho_{ijkl}^0)(\rho_{klij}-\rho^0_{klij}).
\end{equation}

In summary, we propose a new method to compute the entanglement change  in
a dynamical process.We see  the above treatment in Sec.2 and Sec. 3
does not refer to specific entangled systems. This is a desired property as
it makes our measure of entanglement universal. Especially, the results
yielded by present paper can be easily generalized to more than two
subsystems,
this is just the case of many qubits interacting simultaneously with
environment. In addition to the measure stated above,
the quantum relative entropy defined as
$$D(\rho(t)||\rho_A^0(t)\otimes\rho_B^0(t))=Tr[\rho(t)(ln\rho(t)
-ln\rho_A^0(t)
\cdot\rho_B^0(t))]$$ and the Bures metric given by
$$D(\rho(t)||\rho_A^0(t)\otimes\rho_B^0(t))=2-2\sqrt{F(\rho,\rho_A^0\otimes
\rho_B^0)},$$
with $F(\rho,\rho_A^0\otimes\rho_B^0)=[Tr[\sqrt{\rho_B^0\otimes\rho_A^0}\rho
\sqrt{\rho_B^0\otimes\rho_A^0}]^{\frac 1 2 }]^2 $
are other good measures of entanglement. With this modified definitions,
the measures of entanglement can be given in easy way.

\end{document}